\documentstyle[twocolumn]{jpsj}

\def\runtitle{
Josephson current between spin-triplet and spin-singlet
superconductors
} 

\def\runauthor{
Yasumasa {\sc Hasegawa}
 }

\begin{document}
\title {
Josephson Current between Triplet and Singlet
Superconductors
}

\author{ 
Yasumasa {\sc Hasegawa}
}

\inst{
Faculty of Science, Himeji Institute of Technology \\
Kamigori, Akou-gun, Hyogo 678-1297, Japan }

%\recdate{July 28, 1998}
\recdate{ \today}
%\recdate{}

%%%%%%%%%%%%%%_ABSTRACT_%%%%%%%%%%%%%%%%%%%%%%%%%%%
\abst{
The Josephson effect between triplet and singlet
superconductors is studied. 
Josephson current can flow between triplet and singlet
superconductors
due to the spin-orbit coupling in the spin-triplet superconductor
 but it is finite 
 only when triplet superconductor has $L_z=-S_z=\pm  1$,
 where $L_z$ and $S_z$ are the
perpendicular components of orbital angular momentum and spin
angular momentum of the triplet Cooper pairs, respectively.  
The recently observed temperature and orientational
dependence of the critical current
through a Josephson junction between UPt$_3$ and Nb is investigated by 
considering 
a non-unitary triplet state.
}

\kword{
triplet superconductor, odd-parity, spin-orbit coupling
Josephson junction}

%\begin{document}
\sloppy
\maketitle

%%body of paper%%

%%%%%%%%%%%%%%%%%%%%%%%%%%%%%%%%%%%%%%%%%%%%%%%%%%
%\section{Introduction}
Superconductivity in UPt$_3$ attracts a lot of interests.  
The existence of two different superconducting phases 
($A$ and $B$  phases) as well as
 the third phase in the magnetic field  ($C$ phase)
shows that the superconductivity in this system is not a
conventional s-wave spin-singlet superconductivity. 
%The spin-orbit coupling has been thought to be strong in UPt$_3$. 
%There are many experimental evidences that
% spin triplet superconductivity is thought to be realized in
%UPt$_3$.
The possibility of the triplet superconductivity has been 
discussed.\cite{Machida91,Choi91,Sigrist91,Ohmi93}
The anisotropy of $H_{c2}$\cite{Shivaram86} 
is discussed as an evidence for the triplet superconductivity with the
${\mib d}$ vector parallel to the $c$-axis.\cite{Choi91} 
Experimental results of 
the Knight shift observed by the NMR\cite{Kohori87,Tou96,Tou98} and
$\mu$SR\cite{Luke91} 
show the evidence for the triplet superconductivity in
UPt$_3$.\cite{Machida98}
The Knight shift is observed to be independent of temperature
in the superconducting phase for $H > 5$ kOe.\cite{Kohori87,Tou96,Tou98}
It decreases below $T_{\rm c}^+$  
(A and B phases) for $H<5$ kOe when
 $H \parallel \hat{\mib{b}}$ and below $T_{\rm c}^-$ (B phase)
 for $H<2.3$ kOe when  $H \parallel \hat{\mib c}$.\cite{Tou98}
The spin susceptibility for the spin-singlet superconductivity
decreases as temperature becomes low, while that for the spin-triplet
 superconductivity is independent of temperature if the magnetic 
field is perpendicular to the ${\mib d}$ vector. 
Therefore, the magnetic-field and temperature dependences of the
Knight sift of UPt$_3$ can be understood by 
the spin triplet states. 
The A and B phases are identified with
${\mib d}({\mib k}) = d_b({\mib k})\hat{\mib b}$  and  
 ${\mib d}({\mib k}) = d_b({\mib k}) \hat{\mib b} + {\rm i}
d_c({\mib k}) \hat{\mib c}$, respectively.\cite{Tou98,Ohmi96,Machida98}

Since the Josephson effect is controlled by both the amplitude and the
phase of the order parameters, it is a powerful tool to identify the
symmetry of the order parameters. 
%In the High $T_{\rm c}$ materials the
%d-wave spin singlet superconductivity is confirmed by the experiments
%of the Josephson effect.\cite{Wolleman93,Kouznetsov97}   
Recently, Sumiyama {\it et al.}\cite{Sumiyama98}
 have observed the Josephson critical
current between a single crystal UPt$_3$ and 
an s-wave superconductor Nb.
They also  observed the Shapiro steps of the  $\hbar \omega / ( 2 e)$.
If the Josephson current is caused by the higher-order tunneling or
the proximity induced tunneling, the steps of  $\hbar \omega / ( 2 n
e)$ with $n \ge 2$ should be observed.\cite{Yip93,Thuneberg88} 
Thus it is concluded that the Josephson current 
between UPt$_3$ and Nb is 
 caused by the ordinary pair
tunneling of the Cooper pairs. 
They found that  the critical current is  very small in the A phase and
it increases steeply as temperature becomes lower than $T_{\rm c}^-$ 
for the current parallel to the $c$-axis.  
The Josephson  current along the
$b$-axis is  observed even above $T_{\rm c}^-$ and the increasing rate
of the critical current as decreasing temperature 
 becomes slower below $T_{\rm c}^-$.

%
%%%revised%%%%
%
The order parameters near the interface may be
 different from those in the bulk. 
Recently, it has been shown that
the localized zero energy state exists at the interface when the order 
parameter change sign through the reflection\cite{Hu94}.
The zero energy state 
has been studied  for the d-wave 
superconductors\cite{Hu94,Tanaka95,Barash96,Alff97} and also for the
triplet superconductor\cite{Yamashiro97}. 
Although 
the zero energy state should have an influence 
on the Josephson current,
most of the essential features of the Josephson current
 can be understood by considering the lowest order
in the tunneling Hamiltonian for the uniform order
parameters in both
singlet and triplet superconductors.

The Josephson current between spin triplet and spin singlet
superconductors
 is forbidden if  tunneling Hamiltonian does not change the 
spin.\cite{Pals77} 
Due to the spin-orbit coupling in UPt$_3$,
the spin-triplet parings are actually the odd-parity parings, i.e. 
a triplet state is formed in the pseudospin, which is a superposition
of the states with different spins. Although the temperature
independent Knight shift for $H>5$ kOe 
suggests a weak spin-orbit coupling in
UPt$_3$, the  spin-orbit coupling cannot be neglected.
Fenton\cite{Fenton85} and
Geshkenbein and Larkin\cite{Geshkenbein86}
 have shown that  the
Josephson effect between triplet and singlet superconductors
can occur due to the spin-orbit coupling. 
The Josephson tunneling between conventional and unconventional
superconductors are also studied by several 
authors\cite{Millis88,Yip90,Yip93,Yip96}, 
but they assumed the unitary states 
for the triplet superconductors.

In this letter we consider a Josephson current between triplet and
singlet superconductors with the spin-orbit coupling in the triplet
superconductor by taking account of the non-unitary triplet
superconductivity. 
We write the wave number, spin and the creation operator
for the triplet (singlet) superconductor as 
 ${\mib k}$, $\mu$ and $a^{\dagger}_{{\mib k},\mu}$ 
(${\mib l}$, $\nu$ and $b^{\dagger}_{{\mib l},\nu}$), respectively. 
Then the tunneling Hamiltonian in the presence of spin-orbit coupling is
given by
\begin{equation}
  {\cal H}_T= \sum_{{\mib k},{\mib l}} \sum_{\mu,\nu}
    \{ T_{\mu,\nu}({\mib k},{\mib l}) a_{{\mib k},\mu}^{\dagger} 
    b_{{\mib l},\nu} + h.c. \} ,
\end{equation}
where
\begin{equation}
  T_{\mu \nu}({\mib k},{\mib l}) = T({\mib k},{\mib l})
  \delta_{\mu \nu} 
  + T' ({\mib k},{\mib l}) (\hat{\mib k}\times \hat{\mib n}) 
  \cdot \vec{\bf \sigma}_{\mu \nu} .
\label{Tmunu}
\end{equation} 
The first term in eq.(\ref{Tmunu}) is the 
spin diagonal matrix element,
 and the second term is the tunneling matrix element
due to the spin-orbit coupling,\cite{Geshkenbein86}
 where $\hat{\mib n}$ is a unit vector normal to the interface,
$\hat{\mib k}={\mib k}/k_F$ (Fermi surface is assumed to be
spherical),
and  $\vec{\bf \sigma}=(\sigma^x, \sigma^y, \sigma^z)$ is 
a vector with Pauli matrices.
In the second order perturbation in ${\cal H}_T$ the Josephson
coupling energy is calculated as
\begin{equation}
  \Delta E = \langle 0 | {\cal H}_T \frac{1}{E_0 - {\cal H}_0}
  {\cal H}_T | 0 \rangle ,
\label{deltaE}
\end{equation}
where ${\cal H}_0$ is a direct sum of the Hamiltonian for two
superconductors and $E_0$ is the ground state energy for ${\cal H}_0$.
The Josephson current is calculated as\cite{Ambegaokar63}
\begin{eqnarray}
  I &=& -e \langle 0|
  \dot{N}_a \frac{1}{E_0 - {\cal H}_0}{\cal H}_T |0 \rangle
\nonumber \\
& & -e  \langle 0|
   {\cal H}_T \frac{1}{E_0 - {\cal H}_0} \dot{N}_a  |0 \rangle,
\end{eqnarray}
where
\begin{equation}
  N_a=\sum_{\mib k} \sum_{\mu} 
    a^{\dagger}_{{\mib k},\mu}  a_{{\mib k},\mu}, 
\end{equation}
and
\begin{eqnarray}
  \dot{N}_a &=& \frac{\rm i}{\hbar} [N_a,{\cal H}_T]
\nonumber \\
 &=& - \frac{\rm i}{\hbar} \sum_{{\mib k},{\mib l}} \sum_{\mu,\nu}
    \{ T_{\mu,\nu}({\mib k},{\mib l}) a_{{\mib k},\mu}^{\dagger} 
    b_{{\mib l},\nu} - h.c. \}.
\end{eqnarray} 
The Bogoliubov transformations for triplet and singlet 
superconductors are written 
as
\begin{eqnarray}
   a_{{\mib k} \mu}&=& \sum_{\mu'} 
    \left(
     u^{(t)}_{{\mib k}\mu \mu'} \alpha_{{\mib k} \mu'} 
   + v^{(t)}_{{\mib k}\mu\mu'}\alpha_{-{\mib k}\mu'}^{\dagger}
      \right) ,
\end{eqnarray}
and
\begin{eqnarray}
  b_{{\mib l} \nu}&=& \sum_{\nu'}
  \left( 
     u^{(s)}_{{\mib l}\nu \nu'} \beta_{{\mib l} \nu'} 
   + v^{(s)}_{{\mib l}\nu\nu'}\beta_{-{\mib l}\nu'}^{\dagger}
  \right) .
\end{eqnarray}
For the singlet superconductor 
 the order parameter is written as
\begin{equation}
  \Delta^{(s)}_{\nu \nu'} ({\mib l})
  =\Psi({\mib l}) \left( {\rm i} \sigma^y \right)_{\nu \nu'} ,
\end{equation}
and the Bogoliubov transformation is given by 
\begin{eqnarray}
  u^{(s)}_{{\mib l}\nu\nu'} &=& \sqrt{\frac{E_{\mib l}+\epsilon_{\mib l}}
  {2 E_{\mib l}}} \delta_{\nu\nu'}
\nonumber \\
  v^{(s)}_{{\mib l}\nu\nu'} &=& \frac 
  {(-{\rm i}\sigma_y)_{\nu\nu'}\Psi({\mib l})}
  {\sqrt{ 2 E_{\mib l}(E_{\mib l}+\epsilon_{\mib l})}},
\end{eqnarray}
where 
\begin{equation}
 E_{\mib l}=\sqrt{\epsilon_{\mib l}^2 + |\Psi({\mib l})|^2 } .
\end{equation}
The order parameter for spin triplet is written as
\begin{equation}
  \Delta^{(t)}_{\mu \mu'}({\bf k}) 
   = \left( {\rm i}(\vec{\bf \sigma}\cdot {\mib d}({\mib k}))
   \sigma^y \right)_{\mu \mu'}.
\label{deltat}
\end{equation}
If ${\mib d}({\mib k}) \times {\mib d}^*({\mib k}) =0$, 
the order parameter given in eq.~(\ref{deltat}) is proportional to 
a unitary matrix and it is called as a unitary state. 
If ${\mib d}({\mib k}) \times {\mib d}^*({\mib k}) \neq 0$, the state
is called non-unitary. 
For the unitary state the Bogoliubov transformation 
is given by\cite{Sigrist91}
\begin{eqnarray}
  u^{(t)}_{{\mib k}\mu\mu'} &=& 
  \sqrt{\frac{E_{\mib k}+\epsilon_{\mib k}} 
    {2E_{\mib k}}} \delta_{\mu,\mu'}
\nonumber \\
  v^{(t)}_{{\mib k}\mu\mu'}&=&\frac{(-{\rm i}(\mib d({\mib k})\cdot
    \vec{\sigma}) \sigma_y)_{\mu\mu'} }
  { \sqrt{2 E_{\mib k}(E_{\mib k}+\epsilon_{\mib k})}}
\end{eqnarray}
where 
\begin{equation}
   E_{\mib k} = \sqrt{\epsilon_{\mib k}^2+|{\mib d}({\mib k})|^2}.
\end{equation}
We get 
\begin{eqnarray}
 \Delta E &=& 2 {\rm Re} \sum_{{\mib k}{\mib l}}
 \frac 
 {T({\mib k},{\mib l}) T'(-{\mib k},-{\mib l}) 
  (\hat{\mib k} \times \hat{\mib n})\cdot {\mib d}^*({\mib k})
 \Psi({\mib l})}
 {E_{\mib k}E_{\mib l}(E_{\mib k}+E_{\mib l})} 
\nonumber \\
\label{energy}
\end{eqnarray}
In the above we have took only
 terms depending on the phase difference of the order 
parameters and  we have
 neglected the phase independent terms which are irrelevant to the
Josephson critical current. 
%Although we have calculated eq.(\ref{energy}) at $T=0$,
%it may be thought as an approximate
% free energy for the Josephson
%junction at finite temperature, 
%if we take ${\mib d}({\mib k})$ and $\Psi({\mib l})$ 
%as a temperature-dependent order parameters.

The Josephson current between the  spin-triplet unitary state 
and spin singlet state is obtained as\cite{Geshkenbein86}
\begin{equation}
  I =\frac{-2e}{\hbar}
 {\rm Im} \sum_{{\mib k}{\mib l}}\frac 
 {T({\mib k},{\mib l}) T'(-{\mib k},-{\mib l}) 
  (\hat{\mib k} \times \hat{\mib n})\cdot {\mib d}^*({\mib k})
 \Psi({\mib l})}
 {E_{\mib k}E_{\mib l}(E_{\mib k}+E_{\mib l})}
\label{unitary}
\end{equation}
From eq.(\ref{unitary}) it is obtained that the Josephson current
is zero if ${\mib d}({\mib k})$ is parallel to ${\mib n}$. Therefore,
if ${\mib d}$ vector is parallel to the $c$-axis, the Josephson
current along the $c$-axis cannot be explained.
%We consider the Josephson effects between UPt$_3$ and the conventional 
%spin-singlet superconductivity.
%For the current parallel to the $c$-axis we take
% $\hat{\mib n} = \hat{\mib c}$.
Even if ${\mib d}({\mib k})$ is not parallel to $\hat{\mib c}$,
the Josephson current is zero for
 the states belonging to the one-dimensional representation 
of the $D_{6h}$ symmetry for the weak spin-orbit coupling case  such as
 ${\mib d}({\mib k}) \propto k_c \hat{\mib b}$ or  ${\mib d}({\mib k})
\propto  k_c (k_b^2+k_b^2) \hat{\mib b}$ 
due to the summation over ${\mib k}$,
since $T({\mib k},{\mib l})$ and $T'({\mib k},{\mib l})$ have
six-fold rotational symmetry for ${\mib k}$ in the $a$-$b$ plane 
in UPt$_3$. These states have finite Josephson current along the $a$-axis.
This can be
understood as follows. Since the total (orbital ($L$) plus 
spin ($S$)) angular
momentum is conserved by the tunneling Hamiltonian, only the Cooper
pairs with angular momentum $L_z=-S_z=\pm 1$ can be
transformed into spin-singlet pairs.  
If ${\mib d}({\mib k}) \parallel \hat{\mib c}$, Cooper pairs are
formed in the states of $S_z=0$, and no Josephson current flows.
Finite Josephson current parallel to the $c$-axis 
is possible for the state ${\mib d}({\mib k}) \times \hat{\mib c} \neq 
 0$
  such as
\begin{enumerate}
\item
${\mib d}({\mib k}) \propto k_c^2(k_a +{\rm i} k_b) \hat{\mib b}$, 
\item
${\mib d}({\mib k}) \propto (k_c^2 -C)(k_a +{\rm i} k_b) \hat{\mib b}$,
\item
${\mib d}({\mib k}) \propto (k_c^2 -C)(k_a)  \hat{\mib b}$,
\end{enumerate}
etc.,
where $C$ is a constant.  
The $c$ component of the angular momentum is 1 for the first and the
second examples and superposition of 1 and $-1$ for the third
example. These states belong to the two-dimensional representation for 
the weak spin-orbit coupling case, since $L_z=1$ and $-1$ make the
basis for the two-dimensional representation. 
The $z$ component of the spin of the Cooper pair is a superposition of 
$1$ and $-1$ in these cases of the unitary states. As a result
Josephson current is not forbidden by the symmetry.
The energy gap is zero at the equator ($k_c=0$) 
and the poles ($k_a=k_b=0$) of the Fermi surface
for the first example, 
at the parallels of latitude ($k_c=\pm \sqrt{C}$)  and poles for the second
example, and  at the parallels of latitude
and lines of longitude  ($k_a=0$) for
the third example. The density of states $N(E)$ 
for  low energy excitations 
 is proportional to $\sqrt{E}$,
$E$, and $E \log (1/E)$ for three examples, respectively.  
The Josephson critical current 
perpendicular to the $c$-axis is zero for the above three cases.

Next we consider  non-unitary states.
For simplicity
 we take $-{\rm i} {\mib d}({\mib k}) \times 
{\mib d}^*({\mib k})  \parallel \hat {\mib z}$, which is equivalent
to   $d_z({\mib k})=0$.
This choice of the
quantization direction for the spin is possible for 
the equal-spin-paring state, i.e., only up spins ($S_z=1$) and down
spins ($S_z=-1$) make
Cooper pairs and no up-down pairs ($S_z=0$) exist 
by suitably chosen quantization
axis in the spin space. 
Note the $z$ direction in the spin space is not necessarily
to be parallel to the $c$-axis in the real space for the weak spin-orbit 
coupling case.
Then the Bogoliubov transformation is given by
\begin{eqnarray}
  u^{(t)}_{{\mib k}\mu\mu'} &=& \sqrt{
  \frac{E_{{\mib k}\mu'}+\epsilon_{\mib k}}
          {2E_{{\mib k}\mu'}}}
   \delta_{\mu\mu'}
\\
  v^{(t)}_{{\mib k}\mu\mu'} &=& 
 \frac{(-{\rm i} ({\mib d}({\mib k})\cdot
        \vec{\sigma})\sigma_y)_{\mu\mu'}}
         {\sqrt{2E_{{\mib k}\mu'}(E_{{\mib k}\mu'}+\epsilon_{\mib k})}}
\end{eqnarray}
where 
\begin{equation}
  E_{{\mib k}\pm} = \sqrt{\epsilon_{\mib k}^2 + |{\mib d}({\mib k})|^2 
  \pm | {\mib d}({\mib k}) \times {\mib d}^*({\mib k})|}.
\end{equation}
We get the Josephson coupling energy for the non-unitary spin-triplet 
 and spin-singlet superconductors as
\begin{eqnarray}
  \Delta E &=&
 2{\rm Re}  \sum_{{\mib k},{\mib l}} 
  T({\mib k},{\mib l}) T'(-{\mib k},-{\mib l})
\nonumber \\
   \times && 
  \left[ \frac{ \left( 
      (\hat{\mib k}\times \hat{\mib n}) \cdot 
      {\mib d}^*({\mib k})+\left( 
      {\rm i}(\hat{\mib k}\times \hat{\mib n})\times
      {\mib d}^*({\mib k}) \right)_z \right) \Psi({\mib l})}
   {2 (E_{{\mib k}+}+E_{\mib l})E_{{\mib k}+} E_{\mib l}} \right.
 \nonumber \\
  + &&  \left.   
    \frac{ \left( 
      (\hat{\mib k}\times \hat{\mib n}) \cdot 
      {\mib d}^*({\mib k})-\left( 
       {\rm i}(\hat{\mib k}\times \hat{\mib n})\times
      {\mib d}^*({\mib k})\right)_z \right) \Psi({\mib l})}
   {2 (E_{{\mib k}-}+E_{\mib l})E_{{\mib k}-} E_{\mib l}} \right].
\nonumber \\
\end{eqnarray}
The Josephson current between non-unitary spin-triplet and spin singlet 
superconductors is obtained as
\begin{eqnarray}
  I &=&
 \frac{-2e}{\hbar} {\rm Im}  \sum_{{\mib k},{\mib l}} 
  T({\mib k},{\mib l}) T'(-{\mib k},-{\mib l})
\nonumber \\
 & &\times
  \left[ \frac{ \left(
      (\hat{\mib k}\times \hat{\mib n}) \cdot 
      {\mib d}^*({\mib k})+\left( 
      {\rm i}(\hat{\mib k}\times \hat{\mib n})\times
      {\mib d}^*({\mib k}) \right)_z \right) \Psi({\mib l})}
   {2 (E_{{\mib k}+}+E_{\mib l})E_{{\mib k}+} E_{\mib l}} \right.
 \nonumber \\
 & &  
  \left. 
+ \frac{ \left(
      (\hat{\mib k}\times \hat{\mib n}) \cdot 
      {\mib d}^*({\mib k})-\left( 
       {\rm i}({\mib k}\times \hat{\mib n})\times
      {\mib d}^*(\hat{\mib k})\right)_z \right) \Psi({\mib l})}
   {2 (E_{{\mib k}-}+E_{\mib l})E_{{\mib k}-} E_{\mib l}} \right] .
\nonumber \\
\label{nonunitaryIc}
\end{eqnarray}
By noting that
\begin{eqnarray}
 &&  (\hat{\mib k} \times \hat{\mib n}) \cdot {\mib d}^*({\mib k}) 
  \pm {\rm i} \left( (\hat{\mib k} \times \hat{\mib n}) \times  
  {\mib d}^*({\mib k}) \right)_z 
\nonumber \\
 &=& [(\hat{\mib k} \times \hat{\mib n})_x  \mp {\rm i} 
      (\hat{\mib k} \times \hat{\mib n})_y ] 
  [d^*_x({\mib k}) \pm {\rm i} d^*_y({\mib k})] ,
\end{eqnarray}
we obtain that
the Josephson current along the $z$-axis is zero
for the states with
 $L_z = S_z = 1$,
which are the case for example in
${\mib d}({\mib k}) \propto
 (k_x+{\rm i}k_y)(\hat{\mib x} + {\rm i}\hat{\mib y})$ or
${\mib d}({\mib k}) \propto
(k_z^2 - C) (k_x+{\rm i}k_y) (\hat{\mib x} + {\rm i}\hat{\mib y})$.
In the non-unitary states with $L_z=-S_z=1$ such as 
$(k_x+{\rm i}k_y)(\hat{\mib x} - {\rm i}\hat{\mib y})$ or 
$(k_z^2 - C) (k_x+{\rm i}k_y) (\hat{\mib x} - {\rm i}\hat{\mib y})$ 
only the electrons with down spin form Cooper pairs  and
the Josephson current can be finite along the $z$-axis.

Here we assume
 the B phase of UPt$_3$  as a non-unitary state with
\begin{equation}
{\mib d}({\mib k}) = d_b({\mib k})\hat{\mib b} 
+ i d_c({\mib k}) \hat{\mib c},
\end{equation}
as proposed to explain  the Knight shift.\cite{Tou98,Machida98}
 We take $\hat{\mib a}$, $\hat{\mib b}$ and $\hat{\mib c}$ 
as $\hat{\mib z}$, $\hat{\mib x}$ and
$\hat{\mib y}$ respectively in order to have $-{\rm i}
 {\mib d}({\mib k}) \times
{\mib d}^*({\mib k}) \parallel \hat{\mib z}$.
We take $d_c({\mib k})/d_a({\mib k})$ to be real and  we get
\begin{equation}
   E_{{\mib k}\pm}= 
   \sqrt{\epsilon_{\mib k}^2 + |d_b({\mib k}) \pm d_c({\mib k})|^2}.
\end{equation}
For the $c$-axis junction we put 
$\hat{\mib n}=\hat{\mib y}=\hat{\mib c}$ 
and  get the Josephson current as
\begin{eqnarray}
  I_c &=& 
  \frac{-2e}{\hbar} {\rm Im} \sum_{{\mib k},{\mib l}} 
  T({\mib k},{\mib l}) T'(-{\mib k},-{\mib l})
\nonumber \\
 & &\times
  \left[ \frac{-k_a (d_b^*({\mib k}) +  d_c^*({\mib k}))
        \Psi({\mib l})}
   {2 (E_{{\mib k}+}+E_{\mib l})E_{{\mib k}+} E_{\mib l}} \right.
 \nonumber \\
 & &  
  \left. 
 + \frac{-k_a (d_b^*({\mib k}) -  d_c^*({\mib k}))
        \Psi({\mib l})}
   {2 (E_{{\mib k}-}+E_{\mib l})E_{{\mib k}-} E_{\mib l}} \right] .
 \nonumber \\
\end{eqnarray}
For the $b$-axis junction ($\hat{\mib n}=\hat{\mib x}=\hat{\mib b}$), 
we get the Josephson current as
\begin{eqnarray}
  I_b &=& 
  \frac{-2e}{\hbar} {\rm Im} \sum_{{\mib k},{\mib l}} 
  T({\mib k},{\mib l}) T'(-{\mib k},-{\mib l})
\nonumber \\
 & &\times
  \left[  \frac{-{\rm i} k_a (d_b^*({\mib k}) +  d_c^*({\mib k}))
        \Psi({\mib l})}
   {2 (E_{{\mib k}+}+E_{\mib l})E_{{\mib k}+} E_{\mib l}} \right.
 \nonumber \\
 & &  
  \left. 
 + \frac{ {\rm i} k_a (d_b^*({\mib k}) -  d_c^*({\mib k}))
        \Psi({\mib l})}
   {2 (E_{{\mib k}-}+E_{\mib l})E_{{\mib k}-} E_{\mib l}} \right] .
 \nonumber \\
\label{Ib}
\end{eqnarray}

The Josephson current for the $a$-axis junction 
($\hat{\mib n}=\hat{\mib z}=\hat{\mib a}$) is given by
\begin{eqnarray}
  I_a &=& 
  \frac{-2e}{\hbar} {\rm Im} \sum_{{\mib k},{\mib l}} 
  T({\mib k},{\mib l}) T'(-{\mib k},-{\mib l})
\nonumber \\
 & &\times
  \left[ \frac{{\rm i} (k_b -{\rm i} k_c)
       (d_b^*({\mib k}) +  d_c^*({\mib k}))
        \Psi({\mib l})}
   {2 (E_{{\mib k}+}+E_{\mib l})E_{{\mib k}+} E_{\mib l}} \right.
 \nonumber \\
 & &  
  \left. 
 + \frac{-{\rm i} (k_b +{\rm i} k_c)
     (d_b^*({\mib k}) -  d_c^*({\mib k}))
        \Psi({\mib l})}
   {2 (E_{{\mib k}-}+E_{\mib l})E_{{\mib k}-} E_{\mib l}} \right] .
 \nonumber \\
\end{eqnarray}
If we take $d_b({\mib k}) \neq 0$ for $T<T_{\rm c}^+$ (A and B phases) and
 $d_c({\mib k}) \neq 0$ for $T<T_{\rm c}^-$ (B phase),  
it will be difficult to explain the experimental 
results that the Josephson critical current is very small in the A phase 
and increases steeply below $T_{\rm c}^-$ for $I \parallel c$ while $I_b$
  increases more slowly below $T_{\rm c}^-$ 
than in the A phase.\cite{Sumiyama98}

On the other hand if we assume $d_c({\mib k}) \propto k_a$,  
$(k_c^2-C)k_a$, $k_a + {\rm i} k_b$, or $(k_c^2-C) (k_a +{\rm i} k_b)$
 becomes finite below $T_{\rm c}^+$ and  $d_b({\mib k})$ 
 becomes finite below $T_{\rm c}^-$ in the same ${\mib k}$ dependence as 
 $d_c({\mib k})$, we get $I_c$
is zero for $T_{\rm c}^- < T < T_{\rm c}^+$ and 
it increases as $d_b^*({\mib k})\Psi({\mib l})$ 
for $T < T_{\rm c}^-$. In this assumption we get $I_b$ is 
finite below $T_{\rm c}^+$. 
Below $T_{\rm c}^-$ $I_b$ is not changed in the linear
order in $|d_b({\mib k})|$ but it changes in the higher order in
$|d_b({\mib k})|$ and  the second  term in eq.(\ref{Ib})
becomes small when $d_b({\mib k}) \approx d_c({\mib k})$. Thus we
expect the slower increase of $I_b$ below $T_{\rm c}^-$.

In conclusion we have shown that the Josephson current between triplet 
and singlet superconductors is allowed if the Cooper pairs in the 
triplet superconductors have a component $L_z=-S_z=\pm 1$, where $z$
is the direction normal to the junction.  
These results are obtained by 
the lowest order perturbation in the tunneling Hamiltonian
at $T=0$, but the extension to a finite temperature is
possible as in the Junctions between singlet 
superconductors.\cite{Ambegaokar63} 
We can understand the temperature and orientational dependence of the
Josephson critical current\cite{Sumiyama98}
 by assuming the non-unitary state in the B phase, which is proposed
to explain  the Knight shift in NMR\cite{Tou98,Machida98},
although
we have used the different ${\mib d}$ vector in the A phase
 from that is used to
explain the Knight shift.

The author would like to thank A. Sumiyama for showing the
experimental data prior to publication and useful discussions.
%%%%%%%%%%%%%%%%%%%%%%%%%%%%%%%%%%%%%%%%


\begin{thebibliography}{99}


\bibitem{Machida91} 
K. Machida and M. Ozaki: Phys. Rev. Lett. {\bf 66} 
(1991) 3293.


\bibitem{Choi91} 
C.H. Choi and J.A. Sauls: Phys. Rev. Lett. {\bf 66} (1991) 484.


\bibitem{Sigrist91} M. Sigrist and K. Ueda: Rev. Mod. Phys. {\bf 63}
(1991) 239.

\bibitem{Ohmi93} T. Ohmi and K. Machida: Phys. Rev. Lett. {\bf 71} 625 
(1993). 


\bibitem{Shivaram86}
B. Shivaram, T. Rosenbaum and D. Hinks: 
Phys. Rev. Lett. {\bf 57} (1986) 1257.


\bibitem{Kohori87} Kohori et al.; J. Phys. Soc. Jpn. {\bf 56} (1987)
2263; J. Magn. Magn. Matter. {\bf 76 \& 77}(1988) 478.


\bibitem{Tou96} H. Tou, Y. Kitaoka, K. Asayama, N. Kimura, Y. Onuki,
E. Yamamoto and K. Maezawa: Phys. Rev. Lett. {\bf 77} (1996) 1374.

\bibitem{Tou98} H. Tou, Y. Kitaoka, K. Ishida, K. Asayama, N. Kimura,
Y. Onuki, E. Yamamoto, Y. Haga and K. Maezawa: Phys. Rev. Lett. {\bf
80} (1998) 3129.


\bibitem{Luke91} G.M. Luke et al.: Phys. Lett. A{\bf 157} (1991) 173;
J. Appl. Phys. {\bf 70} (1991) 5778.


\bibitem{Machida98} K. Machida and T. Ohmi: J. Phys. Soc. Jpn. {\bf
67} (1998) 1122.


\bibitem{Ohmi96} T. Ohmi and K. Machida: J. Phys. Soc. Jpn. {\bf 65}
(1996) 4018.

\bibitem{Sumiyama98}
A. Sumiyama, S. Shibata, Y. Oda, N. Kimura, E. Yamamoto, Y. Haga and
Y. Onuki: preprint.

\bibitem{Yip93} S. Yip: J. Low Temp. Phys. {\bf 91}, 203 (1993).

\bibitem{Thuneberg88}
E.V. Thuneberg and V. Ambegaokar: Phys. Rev. Lett. {\bf 60} 203 (1993).


\bibitem{Hu94}
C.R. Hu, Phys. Rev. Lett. {\bf 72} (1994) 1526.

\bibitem{Tanaka95}
Y. Tanaka and S. Kashiwaya, Phys. Rev. Lett. {\bf 74} (1995) 3451,
 Phys. Rev. B {\bf 53} (1996) R11957, Phys. Rev. B {\bf 56} (1997)
892.

\bibitem{Barash96}
Y.S. Barash, H. Burkhardt and D. Rainer,
Phys. Rev. Lett. {\bf 77} (1996) 4070.

\bibitem{Alff97} 
L. Alff et al.,
Phys. Rev. B {\bf 55} (1997) R14757.

\bibitem{Yamashiro97}
M. Yamashiro et al.
Physica C {\bf 293} (1997) 239.




\bibitem{Pals77} J.A. Pals, W. Van Haeringen and M.H. Van Maaren:
Phys. Rev. B {\bf 15} (1977) 2592.


\bibitem{Fenton85} E.W. Fenton: Solid State Comm. {\bf 54} (1985) 709.


\bibitem{Geshkenbein86} V.B. Geshkenbein and A.I. Larkin: JETP
Lett. {\bf 43} (1986) 395.

\bibitem{Millis88} A. Millis, D. Rainer and J.A. Sauls: 
Phys. Rev. B {\bf 38}, 4504 (1988).

\bibitem{Yip90} S.K. Yip, O.F. De Alcantara Bonfirm and P. Kumar:
Phys. Rev. B {bf 41} 11214 (1990).


\bibitem{Yip96} S.K. Yip, Y.S. Sun and J.A. Sauls: 
Proc. of LT21, (Prague, 1996),
Czech. J. Phys. {\bf 46} Suppl. S2, 557 (1996). 

\bibitem{Ambegaokar63}
V. Ambegaokar and A. Baratoff:
Phys. Rev. Lett. {\bf 10} (1963) 489, {\bf 11} (1963) 104.


\end{thebibliography}
\end{document}